\documentstyle[12pt]{article}
\def\singlespace {\smallskipamount=3.75pt plus1pt minus1pt
                  \medskipamount=7.5pt plus2pt minus2pt
                  \bigskipamount=15pt plus4pt minus4pt
                  \normalbaselineskip=15pt plus0pt minus0pt
                  \normallineskip=1pt
                  \normallineskiplimit=0pt
                  \jot=3.75pt
                  {\def\smallskip {\vskip\smallskipamount}}
                  {\def\medskip   {\vskip\medskipamount}}
                  {\def\bigskip   {\vskip\bigskipamount}}
                  {\setbox\strutbox=\hbox{\vrule
                    height10.5pt depth4.5pt width 0pt}}
                  \parskip 7.5pt
                  \normalbaselines}
\def\middlespace {\smallskipamount=5.825pt plus1.5pt minus1.5pt
                  \medskipamount=11.25pt plus3pt minus3pt
                  \bigskipamount=22.5pt plus6pt minus6pt
                  \normalbaselineskip=22.5pt plus0pt minus0pt
                  \normallineskip=1pt
                  \normallineskiplimit=0pt
                  \jot=5.825pt
                  {\def\smallskip {\vskip\smallskipamount}}
                  {\def\medskip   {\vskip\medskipamount}}
                  {\def\bigskip   {\vskip\bigskipamount}}
                  {\setbox\strutbox=\hbox{\vrule
                    height15.75pt depth6.75pt width 0pt}}
                  \parskip 7.25pt
                  \normalbaselines}
\def\dblspc {\smallskipamount=7.5pt plus2pt minus2pt
                  \medskipamount=15pt plus4pt minus4pt
                  \bigskipamount=30pt plus8pt minus8pt
                  \normalbaselineskip=30pt plus0pt minus0pt
                  \normallineskip=2pt
                  \normallineskiplimit=0pt
                  \jot=7.5pt
                  {\def\smallskip {\vskip\smallskipamount}}
                  {\def\medskip   {\vskip\medskipamount}}
                  {\def\bigskip   {\vskip\bigskipamount}}
                  {\setbox\strutbox=\hbox{\vrule
                    height21.0pt depth9.0pt width 0pt}}
                  \parskip 15.0pt
                  \normalbaselines}

\def\gm{\gamma }
\def\al{\alpha }

\def\be{\begin{equation}}

\def\j-{\J_-}

\def\ee{\end{equation}}

\def\be{\begin{equation}}
\def\ee{\end{equation}}
\def\al{\alpha}
\def\bea{\begin{eqnarray}}
\def\eea{\end{eqnarray}}
\def\bearr{\begin{eqnarray}}
\def\bearrs{\begin{eqnarray*}}
\def\eearr{\end{eqnarray}}
\def\eearrs{\end{eqnarray*}}
\def\barr{\begin{array}}
\def\earr{\end{array}}
\def\p{\partial}

\def\non\non{\nonumber}
\def\nn8{\nonumber\\[15pt]}
\def\l{\left}
\def\r{\right}
\def\un{\underline}

\def\f{\frac}

\begin{document}
\input epsf
\middlespace
\begin{center}
{\Large Baryogenesis from primordial tensor perturbations
}\\[30pt] Subhendra Mohanty, B. Mukhopadhyay and A. R. Prasanna\\
Physical Research Laboratory\\ Ahmedabad 380 009, India\\[40pt]

\un{Abstract}\\

\end{center}
During inflation primordial quantum fluctuations of the spacetime
metric become classical and there is a spontaneous CPT violation
by the spin connection coupling terms of the metric with fermions.
The energy levels of the left and the right chirality neutrinos is
split which gives rise to a net lepton asymmetry at equilibrium. A
net baryon asymmetry of the same magnitude can be generated from
this lepton asymmetry  either by a GUT,  $B-L$ symmetry or by
electroweak sphaleron processes which preserve $B+L$ symmetry.
  If  the amplitude of the primordial
tensor perturbations is of the order of $10^{-6}$ (as is expected
from inflation models) and the lepton/baryon number violating
processes  freeze out at the GUT era $T_d \sim 10^{16} Gev$  then
a baryon number asymmetry of the correct magnitude $10^{-10}$ can
be generated.

\newpage


The successful prediction of light element abundance by the
big-bang nucleosynthesis \cite{bbn} depends crucially on the assumption
that the net baryon number to entropy ratio $\eta \sim (2.6-6.2) \times
10^{-10}$. In order to generate this non-zero baryon number from
a baryon symmetric universe there are three basic requirements
(Sakharov \cite{sakh}): (i) There should be baryon number
violating  processes  in particle interactions.
 (ii)  $C$ and $CP$ violation to ensure that $B$ generating
 processes are more rapid than $\bar B$ generating
processes and (iii) Out of equilibrium conditions: as CPT demands
that $m_b=m_{\bar b}$, the equilibrium phase space
density of particles and antiparticles are the same. To maintain
$n_b \neq n_{\bar b}$ the reactions should freeze out before
particles and anti-particles achieve thermodynamic equilibrium.

In the standard baryogenesis scenarios \cite{kolb} (i) baryon
number violation occurs in Grand Unified Theories (GUT's) where quarks and
leptons belong to  the same irreducible representation of the
gauge group. (ii) $C$ is  violated maximally in the electro-weak
sector and $CP$ violation is introduced by making the coupling
constants of the lepto-quark gauge bosons complex.(iii) Out of
equilibrium condition is achieved by the expansion of the universe
when the reaction rates become lower than the Hubble expansion
rate at some 'freeze-out' temperature. In GUT baryogenesis
scenarios, the decoupling takes place in the GUT era $T_d \sim
10^{16} Gev$ and the correct baryon-asymmetry parameter is
achieved by tuning the $CP$ violation parameter in the gauge
couplings.

In this paper we show that when the quantum fluctuations of the
metric become classical $CP$ and $CPT$ are violated spontaneously
due to the spin-connection couplings of fermions with gravity. In
the presence of baryon/lepton violating GUT processes there is a
net asymmetry generated between neutrinos and anti-neutrinos at {\it
thermodynamic equilibrium}. This lepton-asymmetry goes down with
temperature and  gets frozen-in when the
lepton-number violating GUT processes decouple. Baryon asymmetry
can be  achieved in two ways (i) if the GUT posseses a $B-L$
residual symmetry \cite{gut} (as in $SU(5)$) then the
generation of lepton-asymmetry is accompanied by baryon asymmetry
of the same magnitude; (ii) the electroweak sphaleron processes
\cite{ew} can generate baryon and lepton number violation at the
electroweak scale. Sphaleron processes conserve $B+L$ so a lepton
asymmetry generated in the GUT era can be converted to baryon
asymmetry of the same magnitude.

The magnitude of the baryon/lepton asymmetry  depends upon the
magnitude of the primordial tensor perturbations and the decoupling
temperature 
\be
\eta \simeq {A_\times~T_d \over g_*^{1/2} M_{pl}}. \ee  If the
primordial tensor perturbations have an amplitude $A_\times \sim
10^{-6}$ , the decoupling temperature $T_d \sim 10^{16} Gev$ and the
relativistic degrees of freedom $g_* \sim 10^{2}$ then the baryon to
photon ratio turns $\eta$ turns
out to be of the required magnitude $\eta \sim 10^{-10}$. This amplitude
of primordial tensor perturbations is consistent with observations
of scalar perturbations by COBE and predictions of many inflation
models \cite{tensor}.

 The  general
covariant coupling of spin $1/2$ particles to gravity is given by
the Lagrangian \cite{kibble}
\be
{\cal{L}} = \sqrt{-g} \l( \bar{\psi}i \gm^a D_a \psi - m
\bar{\psi} \psi \r) \label{l1}
 \ee where the covariant derivative
is given by
 \be
 D_a =  \l( \p_a - \f{i}{4}
\omega_{bca} \sigma^{bc} \r), \ee and the spin-connections are \be
\omega_{bca} = e_{b\lambda}\l(\p_a e^\lambda_{\;\;c} +
\Gamma^\lambda_{\gamma \mu} e^\gamma_{c} e^\mu_a \r), \ee This
Lagrangian  is invariant under the local Lorentz transformation of
the vierbein $e^a_\mu$ and the spinor fields: $e^a_\mu(x)
\rightarrow \Lambda^a_b(x) e^b_\mu (x)$ and the 
$\psi(x) \rightarrow \exp(i \epsilon_{a b}(x) \sigma^{a b})
\psi(x)$. Here $ \sigma^{bc} = \f{i}{2} [\gm^b, \; \gm^c ]$
are generators of tangent space Lorentz transformation (
   $a,b,c$ etc. denote the inertial frame  "flat space"
indices and $\al$, $\beta$, $\gamma$ etc. are the coordinate frame
"curved space" indices such that $e^\mu_a e^{\nu a}= g^{\mu \nu},
e^{a \mu}e^b_\mu= \eta^{ab}, \{\gamma^a, \gamma^b \} = 2 \eta^{a
b}$ where $\eta^{a b}$ represents the inertial frame Minkowski
metric, and $g_{\mu \nu}$ is the curved space metric).

 The spin-connection term in the Dirac equation is a product of
three Dirac matrices. By making use of the identity $ \gamma^a
\gamma^b \gamma^c = \eta^{a b} \gamma^c + \eta^{b c}\gamma^a -
\eta^{a c}\gamma^b - i \epsilon^{a b c d} \gamma_5 \gamma_d $ the
spin connection term can  be reduced to a vector $A^a \gamma_a$
and an axial vector $  B^d \gamma_5 \gamma_d $ coupling. The vector
term
turns out to be anti-hermitian and disappears when the hermetian
conjugate part  is added to the lagrangian (\ref{l1}). The surviving
interaction term which describes the spin-connection coupling of
fermions can be written as a axial-vector
 \bea
 {\cal L}&=& det(e)~\bar \psi \l(~ i \gamma^a \partial_a~ -~m ~-~  \gamma_5 \gamma_d
 B^d~ \r) \psi ,
\nonumber\\
 B^d&=&\epsilon^{abcd} e_{b \lambda}  \l(\partial_a
e^{\lambda}_c + \Gamma^\lambda_{\alpha \mu} e^\alpha_c e^\mu_a \r)
\label{LI}
 \eea
In a local inertial frame of the fermion, the effect of a
gravitational field appears solely as a axial-vector interaction
term shown in (\ref{LI}). We can now calculate the four vector $B^d$
for a perturbed Robertson-Walker universe.

 The general form of perturbations on a flat Robertson-Walker
  expanding universe can be written as \cite{bert}
\be
ds^2= a(\tau)^2 \l[(1+2 \phi) d\tau^2 - \omega_i dx^i d\tau -
\l((1+ 2\psi) \delta_{ij} + h_{ij}\r) dx^i dx^j \r] \ee where
$\phi$ and $\psi$ are scalar , $\omega_i$ are vector and $h_{ij}$
are the tensor fluctuations of the metric. Of the ten degrees of
freedom in the metric perturbations only six are independent and
the remaining four can be set to zero by suitable gauge choice.
For our application we need only the tensor perturbations and we
choose the transverse-traceless (TT) gauge $h^i _i=0 , \partial^i
h_{ij}=0$ for the tensor perturbations. In the TT gauge the
perturbed Robertson-Walker can be expressed as
 \bea
g_{\mu \nu} = a(\tau)^2 \pmatrix{
  1+ 2\phi& -\omega_1 & -\omega_2 & -\omega_3 \cr
  -\omega_1 & -(1+ 2\psi)+h_+  &  h_\times & 0\cr
  -\omega_2 & h_\times & -(1+ 2 \psi)-h_+ & 0 \cr
  -\omega_3& 0& 0& -(1+2 \psi)}
\eea An orthogonal set of vierbiens $e^a_\mu$ for this metric is
given by \bea e_{\mu}^a  = a(\tau) \pmatrix{
  1+ \phi& -\omega_1 & -\omega_2 & -\omega_3 \cr
  0 & -(1+ \psi)+h_+/2  &  h_\times & 0\cr
  0 & 0 & -(1+  \psi)-h_+/2 & 0 \cr
  0& 0& 0& -(1+ \psi)}
  \label{vier}
\eea Using the vierbiens \ref{vier}  the expression for the
components of the four vector field $B^d$ (\ref{LI}) is given by
\bea
 B^0&=& \partial_3 h_\times \nonumber\\ B^1 &=& (\bigtriangledown
\times \vec \omega)^1 \nonumber\\ B^2 &=& (\bigtriangledown \times
\vec \omega)^2 \nonumber\\ B^3 &=&  (\bigtriangledown \times \vec
\omega)^3 + \partial_\tau h_\times \label{B}
 \eea

 The fermion bilinear term $\bar \psi \gamma_5 \gamma_a \psi $
is odd under $CPT$ transformation. When one treats $B^a$ as a
background field  then the interaction term in (\ref{LI})
explicitly violates $CPT$.
 When the primordial metric
fluctuations become classical i.e  there is no back-reaction of
the micro-physics involving the fermions on the metric and $B^a$
is considered as a fixed external field then CPT is violated
spontaneously.

What is important for our application is that the axial
interaction term has different signs for left and right chiral
fields. The axial coupling term for particles $\psi$ and
anti-particles $\psi^c$ may be expressed as \bea  \bar \psi
\gamma_a \gamma_5  \psi &=& \bar \psi_R \gamma_a \psi_R -\bar
\psi_L \gamma_a \psi_L  \label{part}\\ \bar \psi^c \gamma_a
\gamma_5 \psi^c &=& (\bar \psi^c)_R \gamma_a (\psi^c)_R -(\bar
\psi^c)_L \gamma_a (\psi^c)_L  \label{antipart} \eea
  In the standard model neutrinos have left chirality and anti-neutrinos
  have only right chirality so the first term in the neutrino
  coupling (\ref{part}) and the second term in the anti-neutrino
  coupling (\ref{antipart}) are not present. The spin-connection
  interaction has opposite signs for the  (left-handed) neutrinos
and (right-handed) anti-neutrinos. The presence of non-zero neutrino mass
does not affect our analysis as long as $m_\nu << M_{GUT} \sim 10^{16}
Gev$.

 The dispersion relation
of left and right chirality fields are given by
\be
 (p_a \pm B_a)^2=m^2
 \ee
 (here and in the following the upper sign corresponds to $\psi_L$ and the lower sign
 is for $\psi_R$).
Keeping terms linear in the perturbations $B^a$ the free particle
energy of the left  and right  chirality states is
\be
E_{L,R}(p) =(|\vec p| + {m^2\over 2 |\vec p|} ) \mp \l(B_0 -{\vec
p \cdot \vec B \over(|\vec p| } \r) \label{energy} \ee In the
Standard model the left chirality neutrinos $\nu_L$ carry lepton
number $+1$ and right chirality neutrinos $\nu_R$ are assigned
lepton number $(-1)$. In the presence of non-zero metric
fluctuations , there is a split in energy levels of $\nu_{L,R}$
given by (\ref{energy}). If there are GUT processes that violate
lepton number freely above some decoupling temperature $T_d$ then
the equilibrium value of lepton asymmetry generated for all
$T>T_d$ will be
\be
n(\nu_L) -n(\nu_R)= {g \over 2 \pi^2} \int dp^3  \left[ \{1+ exp
~( {E_{L}(p) \over T}) \}^{-1} -\{1+ exp~({E_{R}(p) \over
T})\}^{-1} \right] \label{therm1}
 \ee
 In the ultra-relativistic regime $|\vec p| >> m_\nu $ and
 assuming  that $B_0 << T $ (which is valid at all temperatures
below $ M_{Pl}=1.22\times 10^{19} GeV$ ) the expression
(\ref{therm1}) for lepton asymmetry  reduces to
\be
\Delta n_L = {g T^3 \over 6 }  \l({B_0 \over T}\r)
 \ee
 
 The dependence on $\vec B$ drops out in the angular
integration and the lepton asymmetry depends only on the tensor
perturbations through $B^0$. We can write $B_0$ as a product of an
amplitude $A_\times$ times a wavenumber which represents the
length scale over which the metric fluctuations vary.  The Compton
wavelengths of the particles in the GUT era ( $\sim
 (10^{16} GeV)^{-1}$)  is much smaller than the average
 wavelength of the gravitational waves whose wavenumber $k \sim
 H = 1.66 g_*^{1/2} (T^2/M_{Pl})$.  The gravitational wave
 background can be considered as a constant amplitude field for
 the GUT processes. The mean value of $B_0$ as a function of
 temperature and the primordial tensor wave amplitude $A_\times$
 can be expressed as
 \be
 \langle B_0 \rangle \simeq A_\times H \simeq A_\times \l(1.66~g_*^{1/2}~ {T^2 \over
 M_{Pl}}\r)
 \label{B0}
 \ee
 Here $g_*$ is
the number of relativistic degrees of freedom which for the
standard model  is $106.7$.

The lepton asymmetry (\ref{therm1})  as a function of temperature
can therefore be expressed as (taking $g= 3$ for the three
neutrino flavors)
\be
\Delta n_L(T) \simeq 0.83~ {A_\times~ {g_*}^{1/2} ~ T^4 \over
M_{Pl}}. \ee

   $\Delta n_L (T)$ decreases
with temperature as long as the lepton violating processes are in
thermodynamic equilibrium. When the temperature drops below a
decoupling temperature for the lepton number violating processes,
the value of $\Delta n_L (T<T_d)$  changes only due to expansion
of the scale factor.
 The lepton number to entropy density($s=0.44~ g_*~ T^3$) remains constant after decoupling
 and is given by
\bea \Delta L(T< T_d) \equiv
 {\Delta n_L(T_d) \over s(T_d)}
 &\simeq& 1.89~ {A_\times~T_d  \over g_*^{1/2} M_{Pl}} \nonumber\\
 &\simeq& 1.5 \times 10^{-10} \l({A_\times \over 10^{-6}} \r)  \l({T_d \over 10^{16} Gev}
 \r)
 \l({107\over g_*}\r)^{1/2} .
 \label{etat}
  \eea

  Primordial tensor perturbations (along with the primordial
  scalar perturbations) contribute to the anisotropy of cosmic
  microwave background at large angles. The COBE DMR measurement
  \cite{cobe}
  of temperature anisotropy $\Delta T =30 \mu K$ sets an upper
  limit of $A_\times \leq 10^{-5} $. Tensor perturbations arise in
  inflation models \cite{tensor} in the same way as scalar perturbations of the
  metric \cite{scalar} from quantum fluctuations. The magnitude (and spectrum) of
  the tensor perturbations depend upon the details of inflation
  potential \cite{tensor} , typically they are expected to be an
  order of magnitude smaller in amplitude than scalar
  perturbations (which from COBE are $\sim 10^{-5}$). If one
  chooses
    $A_T\sim 10^{-6}$ and the decoupling temperature of lepton number violating processes
    at the GUT scale $T_d \sim 10^{-16} GeV$ then  a lepton asymmetry of $\Delta L \sim
    10^{-10}$  is generated.
If the GUT has a $B-L$ symmetry then a baryon asymmetry of the
same magnitude and sign is generated as the lepton asymmetry
(\ref{etat}). If the sign of the initial $B_0$ potential
(\ref{B0}) is negative then GUT processes will generate more
anti-neutrinos than neutrinos. In that case spalerons (which
conserve $B+L$) (\cite{ew}) can convert the anti-neutrino excess
to a baryon number asymmetry of the same magnitude and correct
sign.

Spontaneous breaking of CPT to generate baryogenesis at thermal
equilibrium was first introduced in \cite{cohen}. In this model
the baryon number current is coupled to a scalar field $(1/f)
\partial_\mu \phi ~j^\mu_B$. When $\phi$ acquires a vacuum
expactation value ($vev$) and acts like a classical 
field then this term splits the energy levels of baryons and
anti-baryons thereby generating a net baryon asymmetry at thermal
equilibrium. The correct magnitude of $\eta$ depends on the
coupling constant ($1/f$) and the value of the scalar field $vev$.
 A more general class of explicit $CPT$ violating
terms in the lagrangian  which can generate baryon asymmetry have
been studied in \cite{berto}.

In our mechanism the value of $\eta$ is determined by the
decoupling scale (taken as the GUT scale $10^{16} Gev$) and the
amplitude of tensor perturbations ($\sim 10^{-6}$) which is
consistent with COBE observations \cite{cobe} and inflation models
\cite{tensor}. In future observations of CMB polarizations by the
MAP and PLANK satellites it may be possible to accurately
determine the magnitude of tensor perturbations and provide an
experimental check on our baryogenesis model.


\begin{enumerate}

\bibitem{bbn} B.D.Fields and S.Sarkar , in D.Groom et al,
Eur. Phys. J. {\bf C15} (2000);\\
S.Sarkar, Rept. Prog.Phys. {\bf 59} ,1493 (1996).



\bibitem{sakh} A.D.Sakharov , JETP Lett. {\bf 5}, 24 (1967).

\bibitem{kolb} E.W. Kolb and M.S. Turner ,{\it The
Early Universe} , Addison-Wesley , (1990);\\
 A.D.Dolgov, Phys. Rep. {\bf 222 C}, 309 (1992);\\
 A. Riotto and M.Trodden, , hep-ph 9901362.

 \bibitem{kibble} J.Schwinger, {\it Particles, Sources and Fields},
{\bf Vol. 1}, p 401, Addison-Wesley,(1970).

\bibitem{bert} P.J.E. Peebles, {\it Principles of Physical Cosmology},
Princeton Univ. Press, Pinceton, NJ (1993);\\ 
A.R.Liddle and D.H.Lyth {\it
Cosmological
Inflation and Large-Scale structure}, Cambridge Univ Press, (2000).

\bibitem{tensor} V.A.Rubakov, M.Sazhin and A.Veryaskin , Phys.
Lett. {\bf B 115}, 189 (1989);\\ R.Fabbri and M.Pollock,
Phys.Lett. {\bf B 125}, 445 (1983);\\ L.Abbot and M.Wise,
Nucl.Phys.{\bf B 244},541 (1984);\\ B.Allen, Phys. Rev. {D 37}
(1988); L.P.Grischuk , Phys. Rev. Lett. {\bf 70}, 2371 (1993);\\
M.S.Turner, Phys. Rev. {\bf D48}, 3502
 (1993);\\
 M.S.Turner, M.White, J.E.Lidsey, Phys. Rev. {\bf D48} , 4613
 (1993).

\bibitem{scalar} A.H.Guth and S.Y.Pi, Phys.Rev.Lett. {\bf 49}, 1110
(1982);\\ A.A.Starobinskii, Phys. Lett. {\bf B117}175 (1982);\\
J.M.Bardeen, P.J.Steinhardt and M.S.Turner, Phys. Rev {\bf D 28}, 679
(1983).

\bibitem{cobe} G.Smoot {\it et al.}, Astroph. J. {\bf 396}, L1
(1992);\\ E.L.Wright Astroph J. {\bf 396} , L3 (1992).

\bibitem{gut} J.N.Fry, K.A.Olive and M.S.Turner, Phys.Rev.{\bf
D22}, 2953 (1980);\\ J.A.Harvey, E.W.Kolb, B.B.Reiss and
S.Wolfram, Nucl.Phys. {\bf B201} ,16(1982).

\bibitem{ew} V.A.Kuzmin, V.A.Rubakov and M.E.Shapashnikov,
Phys.Lett.{\bf B 155},36 (1985).

\bibitem{cohen} A.Cohen and D.Kaplan , Phys. Lett. {\bf B 199}, 251
(1987).

\bibitem{berto}   O.Bertolami, D.Colladay, V.A. Kostelecky and R.
Potting , Phys Lett {\bf B 395}, 178 (1997).


\end{enumerate}
\end{document}